\documentclass[12pt,a4paper,dvipdfmx]{article}
\usepackage{graphicx}
\usepackage{amssymb,amsfonts,amsmath}
\usepackage{cite}
\usepackage{bm}

\title{Accelerating coordination in temporal networks by engineering the link order}
\bigskip
\author{Naoki Masuda${}^{1}$
\ \\
\ \\
${}^{1}$
Department of Engineering Mathematics, University of Bristol, Bristol, UK.\\
\ naoki.masuda@bristol.ac.uk}

\begin{document}
\setlength{\baselineskip}{24pt}

\maketitle

%\keywords{Keyword1, Keyword2, Keyword3}

\section*{Abstract} % <= 200 words
Social dynamics on a network may be accelerated or decelerated depending on which pairs of individuals in the network communicate early and which pairs do later. The order with which the links in a given network are sequentially used, which we call the link order, may be a strong determinant of dynamical behaviour on networks, potentially adding a new dimension to effects of temporal networks relative to static networks. Here we study the effect of the link order on linear coordination (i.e., synchronisation) dynamics. We show that the coordination speed considerably depends on specific orders of links. In addition, applying each single link for a long time to ensure strong pairwise coordination before moving to a next pair of individuals does not often enhance coordination of the entire network. We also implement a simple greedy algorithm to optimise the link order in favour of fast coordination.


\section*{Introduction}

\begin{figure}[ht]
\centering
\includegraphics[width=8cm]{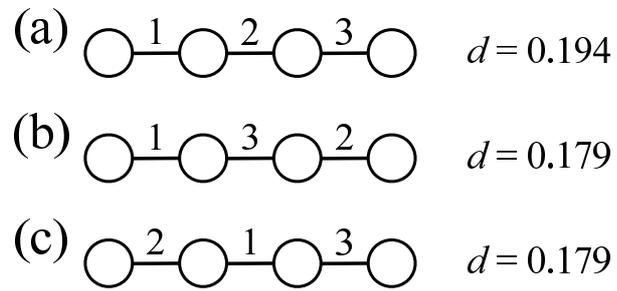}
\caption{The three possible link orders for the chain with $N=4$ nodes. The numbers attached to the links indicate the order with which the link is used. A large value of $d$ implies that coordination occurs rapidly; $d$ is defined in the Results section. The $d$ values are calculated for $\tau=1$, where $\tau$ is the length of time for which each link is applied.}
\label{fig:schem chain N=4}
\end{figure}

Consider a contact network composed of four persons shown in Fig.~\ref{fig:schem chain N=4}. A node shown as a circle represents an individual. A link connecting two nodes represents a dyadic relationship. Suppose that you are in a managerial position to urge them to communicate with each other to induce coordination among them as soon as possible. Because communication is generally costly, you may be interested in making the number or total time of communication small. The four individuals are assumed to be too conservative or time-restricted to change the network structure themselves, such that they would only discuss the matter with their extant neighbours. In addition, pairwise communication events may be a main method to exchange information between individuals due to a social norm, the capacity of each individual or the property of the matter to be discussed. In this situation, which pair of individuals should initiate discussion first? Does coordination take place faster if you force them to discuss starting from the leftmost pair to the rightmost pair along the chain (Fig.~\ref{fig:schem chain N=4}(a); the number on the link represents the order of pairwise interaction). Alternatively, is it better for the two individuals in the middle to communicate last and after the other two pairs (Fig.~\ref{fig:schem chain N=4}(b)), or vice versa (Fig.~\ref{fig:schem chain N=4}(c))?

Motivated by this fictive example, in the present study we ask how we can possibly accelerate coordination in a given social network by engineering the order of links to be used. For example, the chain network shown in Fig.~\ref{fig:schem chain N=4} has four nodes and three links. Thanks to symmetry, the three orders with which to use all links just once shown in the figure exhaust all possibilities. The order of link usage, which we call the link order, may impact the speed of coordination among the four nodes (and it in fact does). Specificity of the link order generally influences dynamics occurring on temporal networks \cite{HolmeSaramaki2012PhysRep,HolmeSaramaki2013book_Springer,Holme2015EurPhysJB}. We examine linear diffusion on networks under link switching dynamics and quantify how different link orders yield different levels of coordination at a final time. We also propose a simple greedy algorithm to accelerate coordination.

The effect of the link order on collective dynamics or performance of networks has been examined in at least two fields. First, in the study of cellular automata including random Boolean networks, different methods to update the states of cells, such as synchronous updating (i.e., update all cells simultaneously) versus asynchronous updating (i.e., update cells one by one) and deterministic versus stochastic updating, have been shown to influence dynamics \cite{Gershenson2002ALIFE8,Cornforth2005PhysicaD,Bandini2012NatComput}. However, specific link orders do not seem to be of a primary question in this field. Second, scheduling of link orders has been formulated and optimised via mathematical programming techniques in the context of wireless networks, where links are used for transmission and conventionally interfere with each other \cite{Hajek1988IEEETransInfoTh,Manweiler2012IEEETransNetw}. However, the objective functions and constraints of these models are specific to wireless communications.

\section*{Results}

Consider a static and unweighted network having $N$ nodes and $M$ links. We allow multiedges, which are counted as distinct links. We assume that each node carries continuous state $x_i$ ($1\le i\le N$) that varies over time according to linear diffusive dynamics.
 We sequentially apply links, say, $(i, j)$, to the network to induce linear diffusive dynamics with a coupling strength of unity, corresponding to a pairwise conversation event towards coordination between the $i$th and $j$th nodes. The duration of a link is denoted by $\tau$ and takes a common value for all links. The states of the other $N-2$ nodes are unchanged in this period. The application of a single link shrinks the distance between $x_i$ and $x_j$ by a factor of $e^{-2\tau}$. In other words, the original $x_i$ and $x_j$ are mapped to
 $\left[(1+e^{-2\tau})x_i\right]/2 + \left[(1-e^{-2\tau})x_j\right]/2$ and $\left[(1-e^{-2\tau})x_i\right]/2 + \left[(1+e^{-2\tau})x_j\right]/2$, respectively (Eqs.~\eqref{eq:x_i(tau)} and \eqref{eq:x_j(tau)}).
Then, we switch to a next link, which is applied for time $\tau$. This procedure is repeated until all links are used exactly once.
The model can be interpreted as the bounded confidence model \cite{Deffuant2000AdvCompSyst} on networks in which the interaction threshold is equal to zero (i.e., the two nodes interact regardless of the distance between $x_i$ and $x_j$).

The dynamics depend on the link order. There are $M!$ link orders.
The actual number of link orders is usually much smaller than $M!$ for at least three reasons. 
First, multiedges introduce obvious redundancy in the count of link orders.
Second, symmetry in the network structure reduces the effective number of link orders. For example, the chain network shown in Fig.~\ref{fig:schem chain N=4} has three distinct link orders, whereas $M! = 3! = 6$. Third, swapping the order of the two links that are subsequently applied does not affect the dynamics afterwards if the corresponding single-link Laplacians commute. In undirected networks, this condition is met if and only if the two links that are subsequently applied are the same (i.e., multiedges) or they do not share a node (see Methods).

\subsection*{Measure of the speed of coordination}

We introduce $d$, which quantifies the level of coordination when all links have been applied just once, i.e., at time $t=M\tau$. We define $d$ as the normalised mean distance between a pair of nodes when the initial state of each node, $x_i(0)$ ($1\le i\le N$), obeys an independent and identical normal distribution with mean zero and standard deviation $\sigma$. 
A practical sufficient condition for full coordination as $t\to\infty$ is the connectedness of 
the temporal network aggregated over time interval $[t_i, t_{i+1})$ for each $i$, where 
$t_i$ can be selected arbitrarily \cite{Jadbabaie2003IEEEAutoControl,Moreau2005IEEEAutoControl,RenBeard2005IEEEAutoControl}.
We do not consider asymptotic relaxation time, which is more commonly studied, because the present model is motivated by social settings in which pairwise interaction, i.e., a link, is considered to be costly and would not be used infinitely many times.

Denote the state vector by $\bm x(t)\equiv (x_1(t)\; \cdots\; x_N(t))^{\top}$, where $\top$ represents the transposition.
We calculate $d$ in terms of matrix $T$ that maps the initial state vector $\bm x(0)$ to the final state vector $\bm x(M\tau)$, where $T$ is explicitly given by Eq.~\eqref{eq:T def}. The mean square distance between the states of two nodes is initially equal to $2\sigma^2$. The mean square distance at $t=M\tau$ averaged over all node pairs and normalised by the value at $t=0$ is given by
\begin{align}
d \equiv& \frac{1}{2\sigma^2}E\left\{\sum_{i=1}^N \sum_{j=1}^{i-1} \frac{2}{N(N-1)}\left[x_i(t) - x_j(t)\right]^2\right\}\notag\\
=& \frac{1}{N(N-1)\sigma^2}E\left\{\sum_{i=1}^N \sum_{j=1}^{i-1}\left[\sum_{\ell=1}^N T_{i\ell}x_{\ell}(0) - \sum_{\ell=1}^N T_{j\ell} x_{\ell}(0)\right]^2\right\}\notag\\
=& \frac{1}{N(N-1)} \sum_{i=1}^N \sum_{j=1}^{i-1} \sum_{\ell=1}^N (T_{i\ell}-T_{j\ell})^2\notag\\
=& \frac{1}{N(N-1)} \sum_{i=1}^N \sum_{j=1}^{i-1} \left\| \bm b_i - \bm b_j\right\|^2,
\label{eq:d def}
\end{align}
where $E$ denotes the expectation, $\bm b_i = (T_{i1}, \ldots, T_{iN})$ represents the $i$th row of $T$, $\left\| \cdot \right\|$ denotes the $L^2$-norm of the vector, and we have used the independence between $x_i(0)$ and $x_j(0)$ ($i\neq j$) to derive the second last equality in Eq.~\eqref{eq:d def}.
Equation~\eqref{eq:d def} indicates that $d$ is small if the rows of $T$ are close to each other in terms of the $L^2$-norm. The $d$ value depends on $\tau$, i.e., how long each link is applied. The $d$ values for the three link orders for the network shown in Fig.~\ref{fig:schem chain N=4} are presented in the figure with $\tau=1$.
In the following analysis, we discuss the speed of coordination in terms of $d$ unless otherwise stated.

% We can avoid the vector subtraction to rewrite Eq.~\eqref{eq:d def} as
% \begin{equation}
% d = \frac{1}{N} \sum_{i=1}^N \bm \left\| b_i \right\| ^2 - \frac{1}{N(N-1)}\sum_{i=1}^N \sum_{j=1}^{i-1} \bm b_i \cdot \bm b_j,
% \label{eq:d succinct}
% \end{equation}
% where $\bm b_i \cdot \bm b_j$ is the inner product of the two vectors. However, Eq.~\eqref{eq:d succinct} is susceptible to numerical error more than Eq.~\eqref{eq:d def} is. Therefore, we calculated $d$ using Eq.~\eqref{eq:d def}.

\subsection*{Real temporal networks}

We start numerical analysis with two real temporal networks obtained from human interaction data.
The first data set was obtained from the SocioPatterns Project and was recorded from participants in a conference \cite{Isella2011JTheorBiol}.
Although the original data have $113$ nodes and $20818$ links, we have excluded one node and the two links incident to the deleted node because the speed of coordination is very sensitive to the presence or absence of this node \cite{Masuda2013PhysRevLett}. The reduced network has $N=112$ and $M=20816$ links. We call this data set the conference data set. The second data set was obtained from the Reality Mining Project and was recorded from
students, staff and faculty members at the Massachusetts Institute of Technology\cite{Eagle2006PersUbiquitComput}.
Although the original network contains 106 nodes,
we use the largest connected component containing $N=104$ nodes and 
$M=782682$ links. Both networks contain multiedges. We respect the link orders in the original data sets and calculate $d$. Because of the finite time resolution in the recording, there are often more than one links appearing in the same time window. In this case, we follow the link order as dictated in the original data.

\begin{figure}[ht]
\centering
\includegraphics[width=16cm]{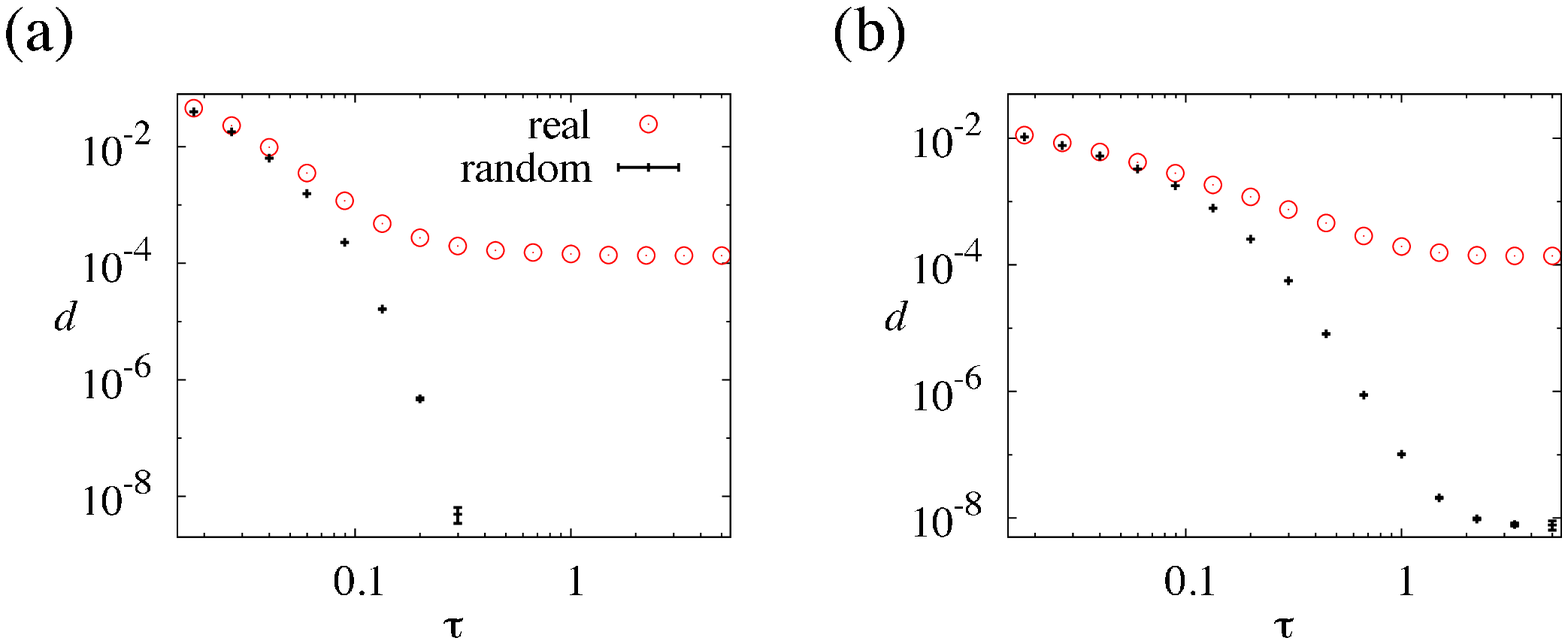}
\caption{Speed of coordination, $d$, for the real temporal networks and random link orders. The error bar represents the average $\pm$ standard deviation. (a) Conference data set. (b) Reality Mining data set.}
\label{fig:temporal}
\end{figure}

The $d$ values are plotted against $\tau$ for the conference and Reality Mining data sets by the circles in Figs.~\ref{fig:temporal}(a) and \ref{fig:temporal}(b), respectively.
The average and standard deviation of $d$ on the basis of $10^3$ randomly generated link orders while the structure of the aggregate (i.e., static) network is kept intact are shown by the error bars in Fig.~\ref{fig:temporal}.
The figure indicates that coordination occurs on real temporal networks much more slowly than for typical random link orders. This result is consistent with our previous results in which we looked at the spectral gap, an alternative measure of coordination, for the same data sets \cite{Masuda2013PhysRevLett}.
The difference between the real temporal networks and random link orders is more significant for larger values of $\tau$. Real temporal networks are slow to coordinate possibly because of temporally correlated appearance of links \cite{Masuda2015IFAC}.
In addition, coordination is not enhanced as $\tau$ increases when $\tau$ is large. 

\subsection*{Small networks}

Small networks allow us to enumerate the link orders and compare the performance of them. In this section, we consider three small networks with $M=10$ links.

\begin{figure}[ht]
\centering
\includegraphics[width=16cm]{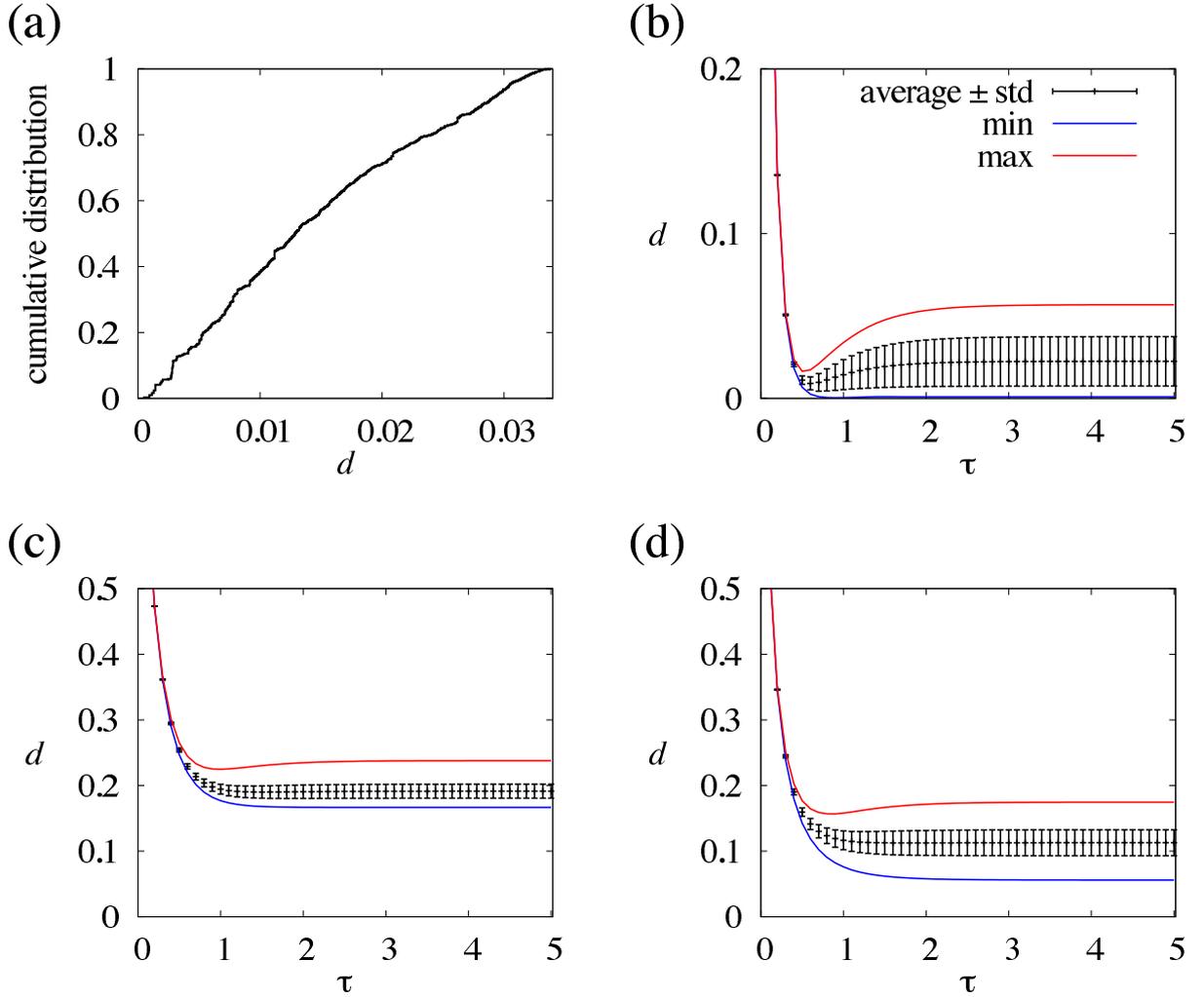}
\caption{Distribution of $d$ for small networks.
(a) Cumulative distribution for all possible link orders for the complete graph with $N=5$ nodes. We set $\tau=1$. (b) Statistics of $d$ for the complete graph with $N=5$ nodes on the basis of all link orders. The error bar indicates the average $\pm$ standard deviation. The minimum and maximum values of $d$ for each $\tau$ are shown by the curves. (c) Statistics of $d$ for the cycle with $N=10$ nodes. (d) Statistics of $d$ for the random graph with $N=7$ nodes and $M=10$ links.}
\label{fig:small}
\end{figure}

We first consider the complete graph, i.e., all node pairs are adjacent by a link. When $N=5$, there are $M=10$ links and $M! = 3628800$ link orders. We set $\tau=1$ and calculate $d$ for all the possible link orders. The cumulative distribution of $d$ is shown in Fig.~\ref{fig:small}(a). The figure indicates that the link order, in combination with the normally distributed initial states, affects $d$ despite that the underlying network is structureless. The value of $d$ is more than 20 times different between the fastest and slowest link orders. Furthermore, both approximately fastest and approximately slowest cases are attained by some fractions of link orders, i.e., not only by exponentially rare link orders.

The average, standard deviation, minimum and maximum of $d$ calculated on the basis of all $M!$ link orders are shown in Fig.~\ref{fig:small}(b) for a range of $\tau$. The error bar indicates the average $\pm$ standard deviation. First, except when $\tau$ is small, $d$ is substantially different between the fastest and slowest link orders, which extends the observation made with Fig.~\ref{fig:small}(a). Second, a large value of $\tau$ indicates that each link is applied for a long time, which might lead one to suspect that a large $\tau$ value improves coordination. However, this is not the case. $d$ decreases as $\tau$ increases only when $\tau$ is small. The fastest coordination is realised between $\tau=0.5$ and $\tau=0.7$ depending on either the average, minimum or maximum of $d$ is considered. When $\tau$ is large, $d$ increases as $\tau$ increases, deteriorating coordination. Anecdotally, forcing each pair of individuals to have long discussion to ensure strong pairwise consensus does not necessarily facilitate coordination at the network level.

The average, standard deviation, minimum and maximum of $d$ for all link orders for the cycle with $N=10$ nodes are shown in Fig.~\ref{fig:small}(c). The average of $d$ is the smallest at $\tau\approx 1.5$. The $d$ value beyond $\tau\approx 1.5$ increases as $\tau$ increases just slightly for this network. The difference between the minimal and maximal $d$ values is smaller than in the case of the complete graph. Nevertheless, the results are qualitatively the same as those for the complete graph. The results are similar for a network with $N=7$ nodes and $M=10$ links generated from the Erd\H{o}s-R\'{e}nyi random graph with the probability of link between a pair of nodes equal to $0.4$
(Fig.~\ref{fig:small}(d)).

\subsection*{Optimising the link order for larger networks}
 
For larger networks, it is prohibitive to determine the best link order. Therefore, we seek to accelerate coordination in terms of $d$ by running a simple greedy algorithm (see Methods). 
For simplicity, we focus on static and unweighted model and empirical networks in this section.

We first analyse the karate-club network in which a node represents a member of the club and a link represents casual interaction between two members. The network has $N=34$ nodes and $M=78$ links \cite{Zachary1977JAnthropolRes}. The link order optimised with $\tau=1$ yields $d=0.0768$. 
For comparison, we also sample $10^3$ random link orders on the same static network and calculate the average and standard deviation of $d$. Randomly sampled link orders yield $d=0.1049\pm 0.0045$ at $\tau=1$ (average $\pm$ standard deviation). This result indicates that the greedy algorithm can find a link order that by far outperforms most link orders.

\begin{figure}[ht]
\centering
\includegraphics[width=16cm]{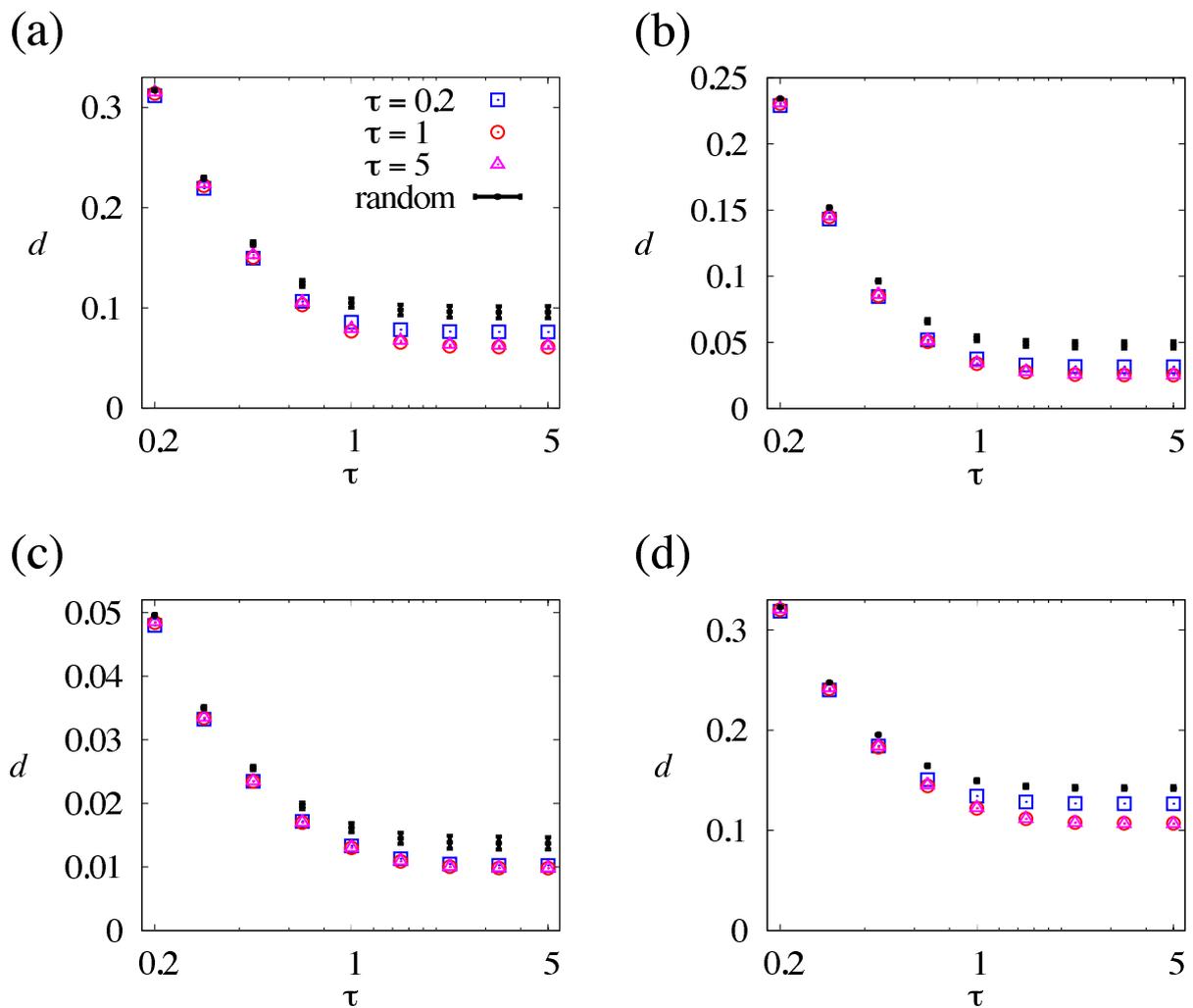}
\caption{Speed of coordination, $d$, for the optimised and random link orders. The symbols correspond to different values of $\tau$ for which the link order is optimised in terms of $d$. The error bar represents the average $\pm$ standard deviation. (a) Karate-club network. (b) BA model. (c) Jazz musician network. (d) Collaboration network.}
\label{fig:d across tau}
\end{figure}

It may be difficult to measure $\tau$ in real settings. Therefore, we assess how the link order optimised for $\tau=1$ performs when the dynamics are actually implemented with different $\tau$ values. For the link order optimised for $\tau=1$, $d$ is plotted against $\tau$ by the circles in Fig.~\ref{fig:d across tau}(a). The average and standard deviation for randomly sampled link orders are shown by the error bars in the same figure. Figure~\ref{fig:d across tau}(a) indicates that the link order optimised for $\tau=1$ also behaves fairly well for other $\tau$ values, in the sense that $d$ is smaller than those for typical random link orders. For the link orders optimised for $\tau=0.2$ and $\tau=5$, $d$ is plotted against $\tau$ by the squares and triangles, respectively, in Fig.~\ref{fig:d across tau}(a). The link orders optimised for these $\tau$ values also yield $d$ values significantly smaller than those for typical random link orders for a range of $\tau$. Therefore, up to our numerical efforts, 
engineering a link order for a certain value of $\tau$ accelerates coordination for a range of $\tau$.

We repeated the same analysis for three other networks.
The results for a heterogeneous network generated by the Barab\'{a}si-Albert preferential attachment model (BA model) having $N=100$ nodes $M=294$ links \cite{Barabasi1999Science} are shown in Fig.~\ref{fig:d across tau}(b). We set the initial network to the triangle and the number of links that each new node possesses to three.
The results for a network of jazz musicians having $N=198$ nodes and $M=2742$ links \cite{Gleiser2003AdvCompSyst} are shown in Fig.~\ref{fig:d across tau}(c). The results for the largest connected component of the collaboration network among major network science researchers~\cite{Newman2006PhysRevE-collabo}, which has $N=379$ and $M=914$ links, are shown in Fig.~\ref{fig:d across tau}(d). The results for the three networks are qualitatively the same as those for the karate-club network.

For all four networks, $d$ for the optimised or random link orders decreases as $\tau$ increases when $\tau$ is small. However, $d$ would not decrease further as $\tau$ increases when $\tau$ is large. Therefore, leaving pairs of individuals for a long time to ensure strong pairwise consensus does not accelerate the formation of consensus at a network level. This result is consistent with that for real temporal networks (Fig.~\ref{fig:temporal}) and small networks (Figs.~\ref{fig:small}(b)--(d)).

\subsection*{Spectral gap}

The relaxation speed of linear diffusive dynamics is usually characterised by the eigenvalue that determines the relaxation time of the dynamics. Because $T$ (Eq.~\eqref{eq:T def}) is a linear map from the initial state vector to the final state vector, the relevant eigenvalue is the second largest eigenvalue of $T$ in terms of the modulus, denoted by $\lambda_2(T)$. The largest eigenvalue of $T$ is always equal to unity, corresponding to the perfectly synchronous mode, or the right eigenvector $(1\; \cdots\; 1)^{\top}$. Here we examine $-\log\left(\left|\lambda_2(T)\right|\right)/\tau$, which corresponds to the spectral gap of the Laplacian dynamics in continuous time under switching dynamics \cite{Masuda2013PhysRevLett}. If the spectral gap calculated for the ordered link sequence $e_1$, $\ldots$, $e_M$ is large, coordination occurs fast when we periodically apply links $e_1$, $\ldots$, $e_M$, $e_1$, $\ldots$ infinitely many times. We prefer $d$ to the spectral gap because pairwise conversations would not probably repeat periodically in real social situations. Nevertheless, in this section we assess whether minimisation of $d$ also enhances the spectral gap.

\begin{figure}[ht]
\centering
\includegraphics[width=16cm]{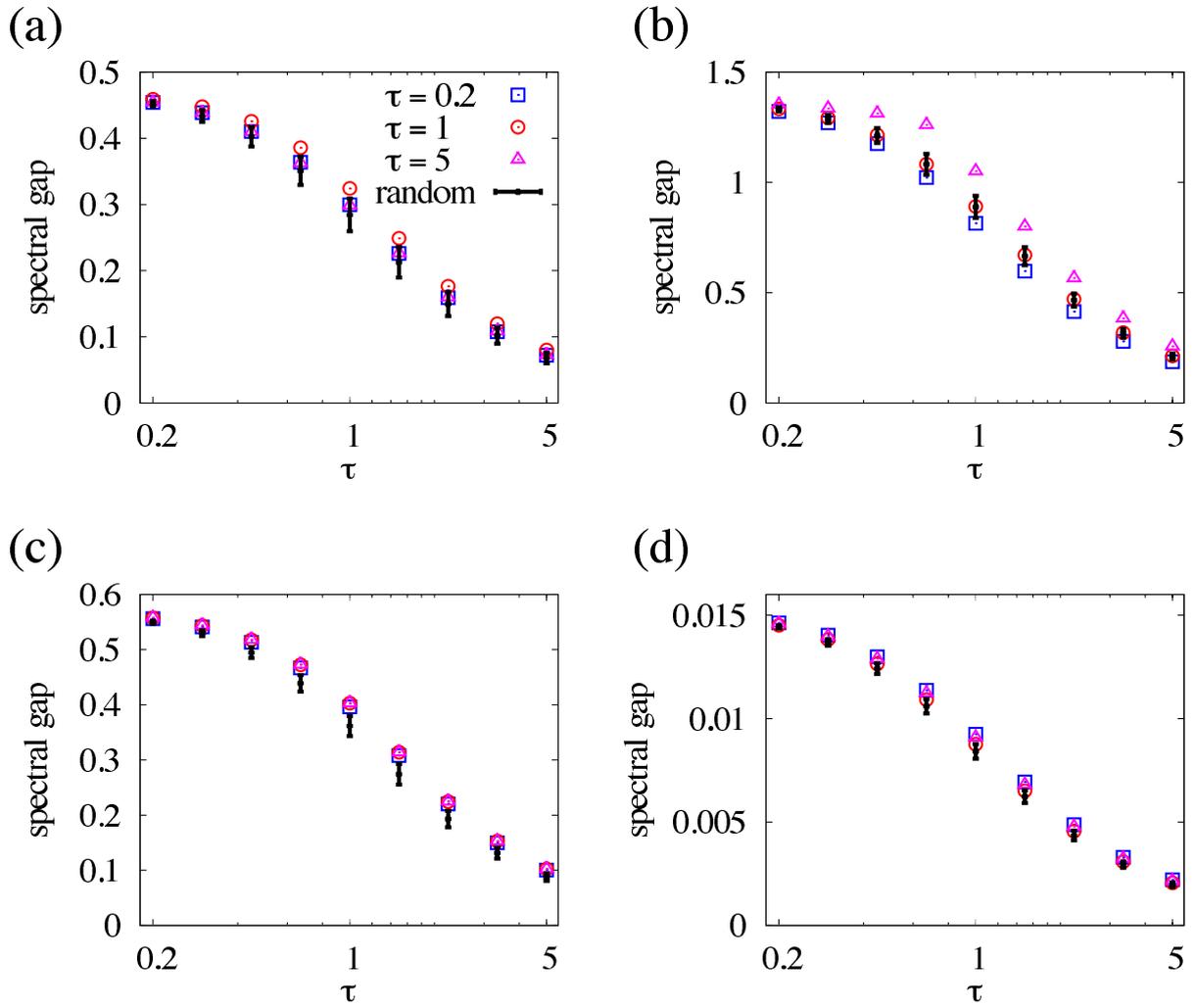}
\caption{Spectral gap for the optimised and random link orders. The symbols represent the results for the three optimised link orders used in Fig.~\ref{fig:d across tau}. (a) Karate-club network. (b) BA model. (c) Jazz musician network. (e) Collaboration network.}
\label{fig:lambda2 across tau}
\end{figure}

We compared the spectral gap for the link orders optimised in terms of $d$ and that for random link orders. The results for the karate-club network, BA model, jazz musician network, and collaboration network are shown in Figs.~\ref{fig:lambda2 across tau}(a), \ref{fig:lambda2 across tau}(b),\ref{fig:lambda2 across tau}(c), and \ref{fig:lambda2 across tau}(d), respectively.
For the karate-club, jazz, and collaboration networks, the link orders optimised in terms of $d$ for the three values of $\tau$ yield larger spectral gaps than the mean value for random link orders, in a range of $\tau$. For example, in the karate-club network (Fig.~\ref{fig:lambda2 across tau}(a)), the link order optimised in terms of $d$ at $\tau=1$ produces the spectral gap values that are larger than the average plus the standard deviation of typical link orders for all the examined values of $\tau$. However, the spectral gap is only slightly larger than in the random case for the link orders optimised with $\tau=0.2$ and $\tau=5$. In the BA model (Fig.~\ref{fig:lambda2 across tau}(b)), the optimisation in terms of $d$ sometimes makes the spectral gap smaller than that for typical link orders. For this network, the link order optimised with $\tau=5$ is the only case in which the spectral gap for the optimised link order is larger than that for typical link orders.

\section*{Discussion}

The present study may be extended in the following aspects.

First, we provided a simple heuristic greedy algorithm to search for a link order to accelerate coordination in terms of $d$. The search space is composed of all permutations on $M$ links. Although multiedges, symmetry in the network structure and commuting single-link Laplacians reduce the search space, the effective search space is generally huge even for a network with a small number of links. The current problem is a permutation-based combinatorial optimisation problem, whose famous examples include the travelling salesman problem and the quadratic assignment problem. Such a problem is typically NP-hard. The optimised solutions found in the present study may be local optimums. However, various approximate algorithms can find nearly optimal solutions for famous permutation-based problems \cite{Ceberio2012ProgArtifIntell}. It may be possible to build heuristic algorithms for the present model that provide better solutions than those obtained in the Results section.

Second, we only allowed isolated pairwise communications in each time period, corresponding to single-link Laplacians. In real situations, communications towards consensus may occur in a group~\cite{Eckmann2004PNAS,Stehle2010PhysRevE,Miritello2011PhysRevE}. It is straightforward to extend the current framework to the case of group conversation. 
For single-link Laplacians, we used Eq.~\eqref{eq:from exp to linear} to transform the linear diffusive dynamics in continuous time to that in discrete time to simplify the computation of $T$, the mapping from the initial state vector to the final state vector. Otherwise, the calculation of $T$ requires computationally burdening matrix exponentials. A group conversation corresponds to a clique in the network. If cliques do not overlap in each snapshot and are of the same size within and across snapshots, a relation similar to
 Eq.~\eqref{eq:from exp to linear} holds true \cite{Masuda2015IFAC}, facilitating the computation of $T$. 
 
Third, we imposed the condition that all links are used exactly once. In real situations, it may be allowed to use the same link multiple times, and some links may not have to be used. 
Burstiness of links and other higher-order temporal and structural correlation as present in empirical data \cite{HolmeSaramaki2012PhysRep} may impact the speed of coordination.
In this situation, a plausible constraint may be to fix the total number of times that single links are applied. Then, the structure of the aggregate network composed of the actually used links varies across link orders. The structure of the static network is generally a strong determinant of the level of coordination (i.e., synchronisation) \cite{Donetti2005PhysRevLett,Almendral2007NewJPhys,Arenas2008PhysRep}. Therefore, the structure of the aggregate network, such as community structure and different levels of heterogeneity in the node's degree, may have a larger impact on $d$ than the link order.

Fourth, we assumed that the initial states of the nodes, $x_i(0)$ ($1\le i\le N$), are independently and identically distributed. If the information about the initial or intermediate states is available, it is probably possible to devise a link order to accelerate coordination. For example, if the states of all nodes are monitored, always forcing the most distant pair of nodes in terms of $x_i(t)$ to communicate may considerably hasten coordination of the entire network.

Fifth, we assumed the linear diffusion dynamics for simplicity. Extening the present framework to nonlinear synchronisation dynamics and other types of dynamics is straightforward. In fact, effects of temporal networks on collective dynamics have been investigated with various models \cite{Holme2015EurPhysJB}. Examples include nonlinear coupled dynamics towards synchronisation \cite{Fujiwara2011PhysRevE}, the voter model \cite{Baxter2011JStatMech,Fernandez-Gracia2011PhysRevE,Takaguchi2011PhysRevE}, the naming game \cite{Baronchelli2012PhysRevE,Maity2012PhysRevE}, the information cascade model \cite{Karimi2013PhysicaA}, dynamics of triangular social balance \cite{Nishi2014EPL}, evolutionary game dynamics \cite{Cardillo2014PhysRevE}, and, above all, epidemic processes \cite{Masuda2013F1000,Pastorsatorras2015RevModPhys}. The link order may add another dimension to the impact of temporal networks on these and other dynamics.

\section*{Methods}

\subsection*{Model}

The model for linear coordination dynamics is the same as that considered in our previous work, in which links are sequentially sampled without replacement \cite{Masuda2013PhysRevLett}.
Assume a static, undirected and unweighted network having $N$ nodes and $M$ links. Multiedges (i.e., multiple links connecting the same pair of nodes) are allowed and are counted as distinct unweighted links. The following formulation can be generalised to the case of directed and weighted networks.

Denote the state of the $i$th node by $x_i\in {\mathbf R}$ ($1\le i\le N$). The network undergoes a sequence of Laplacian dynamics in continuous time in which each link is sequentially applied for time $\tau$. The dynamics for period $0\le t\le \tau$ during which link $(i, j)$ is applied are described by
\begin{align}
\frac{{\rm d}x_i}{{\rm d}t} =& x_j - x_i,\\
\frac{{\rm d}x_j}{{\rm d}t} =& x_i - x_j.
\end{align}
The states of the other $N-2$ nodes are not altered during this period. The states after the application of link $(i, j)$ are equal to $x_i(\tau)$ and $x_j(\tau)$. They are given in terms of the states before the application of the link, i.e., $x_i(0)$ and $x_j(0)$, as follows:
\begin{align}
x_i(\tau) =& \frac{1+e^{-2\tau}}{2} x_i(0) + \frac{1-e^{-2\tau}}{2}x_j(0),
\label{eq:x_i(tau)}\\
x_j(\tau) =& \frac{1-e^{-2\tau}}{2} x_i(0) + \frac{1+e^{-2\tau}}{2}x_j(0).
\label{eq:x_j(tau)}
\end{align}
In short, the application of the link lessens the distance between $x_i$ and $x_j$ by factor of $e^{-2\tau}$.
Then, we apply the next link for $\tau\le t \le 2\tau$ and so forth. The dynamics terminate at $t=M\tau$, when all $M$ links are applied exactly once.

To describe the entire switching dynamics, we introduce the $N\times N$ single-link Laplacian matrix for link $(i, j)$, denoted by $L^{(ij)}$. The matrix is defined by
$(L^{(ij)})_{ii} = (L^{(ij)})_{jj} = 1$ and $(L^{(ij)})_{ij} = (L^{(ij)})_{ji} = -1$. The other elements of $L^{(ij)}$ are equal to zero. Equations~\eqref{eq:x_i(tau)} and \eqref{eq:x_j(tau)} are rewritten as
\begin{equation}
\bm x(\tau) = \exp(-L^{(ij)}\tau)\bm x(0) = \left(I-\epsilon L^{(ij)}\right) \bm x(0),
\label{eq:from exp to linear}
\end{equation}
where $\bm x(t) = (x_1(t), \ldots, x_N(t))^{\top}$, $I$ is the $N\times N$ identity matrix, and
\begin{equation}
\epsilon = \frac{1-e^{-2\tau}}{2}.
\end{equation}
Therefore, for the sequence of links $e_1$, $\ldots$, $e_M$, the final state of the network is determined by
\begin{equation}
\bm x(M\tau) = T \bm x(0).
\end{equation}
where
\begin{equation}
T = (I-\epsilon L^{(e_M)})
(I-\epsilon L^{(e_{M-1})}) \cdots (I-\epsilon L^{(e_1)}).
\label{eq:T def}
\end{equation}
The multiplication of $I-\epsilon L^{(e_m)}$ ($1\le m\le M$) to $\prod_{m^{\prime}=1}^{m-1}(I-\epsilon L^{(e_{m^{\prime}})})$ affects the $i$th and $j$th rows, modifying at most $2N$ elements. Therefore, the computation of $T$ requires $O(NM)$ time. 

\subsection*{Commutator of single-link Laplacians}

Two link orders yield the same $d$ value if $T$ is the same for the two link orders. On the basis of Eq.~\eqref{eq:T def}, a sufficient condition for this is that matrices $I-\epsilon L^{(ij)}$ and $I-\epsilon L^{(k\ell)}$ commute for consecutively used links $(ij)$ and $(k\ell)$. In this case, exchanging the order of $(ij)$ and $(k\ell)$ does not affect the $d$ value. Regardless of the value of $\epsilon$, this condition is equivalent to 
\begin{equation}
[L^{(ij)}, L^{(k\ell)}] \equiv L^{(ij)} L^{(k\ell)} - L^{(k\ell)} L^{(ij)} = 0,
\label{eq:commutator = 0}
\end{equation}
where $[\cdot, \cdot]$ is the Lie bracket, also called the commutator.

Equation~\eqref{eq:commutator = 0} is satisfied when two links $(ij)$ and $(k\ell)$ do not share a node or when they are identical. If the two links share just one node, i.e., $i=k$ and $j\neq \ell$, we obtain
\begin{equation}
[L^{(ij)}, L^{(i\ell)}] = \hat{L}^{(ij\ell)},
\end{equation}
where $N\times N$ matrix $\hat{L}^{(ij\ell)}$ contains three elements equal to $+1$ at $(i,j)$, $(j,\ell)$, $(\ell,i)$, three elements equal to $-1$ at $(i,\ell)$, $(j,i)$, $(\ell,j)$, and zero everywhere else.

Although the main text focuses on undirected networks, the framework is also applicable to directed networks. As in the case of undirected networks, the $d$ value does not change after swapping the order of two directed links that are successively applied if the corresponding single-link Laplacians commute (i.e., the Lie bracket is equal to zero). However, the calculation of the Lie bracket for undirected networks does not generalise to the case of directed networks. We denote by $L^{(\overrightarrow{ij})}$ the $N\times N$ Laplacian matrix for the network composed of just a single directed link from the $i$th node to the $j$th node. In other words, $(L^{(\overrightarrow{ij})})_{ii}=1$, $(L^{(\overrightarrow{ij})})_{ij}=-1$, and all the other elements of $L^{(\overrightarrow{ij})}$ are equal to zero. $L^{(\overrightarrow{ij})}$ is an asymmetric matrix. As in the case of the undirected network, the directed single-link Laplacian matrices commute if the two directed links do not share a node. If the two links share one or two nodes, we obtain
\begin{align}
[L^{(\overrightarrow{ij})}, L^{(\overrightarrow{ik})}] =& L^{(\overrightarrow{ik})} - L^{(\overrightarrow{ji})}\quad (k=j \text{ or } k\neq j),
\label{eq:Lie bracket directed 1}\\
[L^{(\overrightarrow{ij})}, L^{(\overrightarrow{ki})}] =& L^{(\overrightarrow{kj})} - L^{(\overrightarrow{ki})}\quad
(k\neq j),
\label{eq:Lie bracket directed 2}\\
[L^{(\overrightarrow{ij})}, L^{(\overrightarrow{ji})}] =& L^{(\overrightarrow{ij})} - L^{(\overrightarrow{ji})},
\label{eq:Lie bracket directed 3}\\
[L^{(\overrightarrow{ij})}, L^{(\overrightarrow{kj})}] =& 0\quad (k\neq i).
\label{eq:Lie bracket directed 4}
\end{align}
Equations~\eqref{eq:Lie bracket directed 1}--\eqref{eq:Lie bracket directed 4} exhaust all cases in which two links share at least one node.
Because the right-hand sides of Eqs.~\eqref{eq:Lie bracket directed 1}--\eqref{eq:Lie bracket directed 4} are linear sums of single-link Laplacians,
$\{L^{(\overrightarrow{ij})}\}$ ($1\le i, j\le N$) forms a Lie algebra. 
The coefficients of $L^{(\overrightarrow{ij})}$ on the right-hand sides, which are equal to $-1$, $0$, or $1$ in the present case,
are called structure constants.

\subsection*{Greedy algorithm}

The greedy algorithm aims at finding a link order that makes $d$ as small as possible. For a given network and the value of $\tau$, we proceeded as follows. First, we randomly ordered the $M$ links and calculated $d$. Second, we swapped the order of a pair of randomly selected links. The order of the other $M-2$ links was unchanged. Third, we calculated $d$ for the new link order. Fourth, if $d$ decreased by the proposed link swapping, we adopted it. Otherwise, we discarded it. We repeated this procedure $1.5\times 10^5$ times. We verified that $d$ did not notably decrease near the end of the repetition in all runs. 

%\bibliography{../../citations}

\section*{Acknowledgements (not compulsory)}

N.M. acknowledges the support provided through JST, CREST, and JST, ERATO, Kawarabayashi Large Graph Project.

\section*{Author contributions statement}

N.M. conceived the research, conducted the analysis, and wrote the manuscript. 

\section*{Additional information}

\textbf{Competing financial interests:} The author declares no competing financial interests.

\newpage

\clearpage

\clearpage

\clearpage

\clearpage

\clearpage

\end{document}